\documentclass[runningheads]{llncs}
\usepackage[T1]{fontenc}
\usepackage{graphicx}
\usepackage{graphicx}
\usepackage{color,xcolor}
\usepackage{multirow}
\usepackage{booktabs}
\usepackage{amsmath}

\usepackage{amssymb}
\usepackage{url}
\usepackage{algpseudocode}
\usepackage{amsfonts}
\usepackage[ruled,linesnumbered]{algorithm2e}
\usepackage{multicol}     
\usepackage{multirow}
\usepackage[misc]{ifsym}
\usepackage{color}
\usepackage{arydshln}

\begin{document}
\title{Cross Cryptocurrency Relationship Mining for Bitcoin Price Prediction}
%
%
\author{Panpan Li\inst{1,2}    \and
        Shengbo Gong\inst{1,2} \and
        Shaocong Xu\inst{1,2}  \and
        Jiajun Zhou\inst{1,2} \textsuperscript{(\Letter)} \and   \\
        Shanqing Yu\inst{1,2}  \and
        Qi Xuan\inst{1,2}}

\authorrunning{P. Li et al.}
%
\institute{
    Institute of Cyberspace Security, Zhejiang University of Technology, \\Hangzhou 310023, China \\ \and
    College of Information Engineering, Zhejiang University of Technology, \\Hangzhou 310023, China\\ 
    \email{jjzhou@zjut.edu.cn}   
}
\maketitle              
\begin{abstract}
    Blockchain finance has become a part of the world financial system, most typically manifested in the attention to the price of Bitcoin.
    However, a great deal of work is still limited to using technical indicators to capture Bitcoin price fluctuation, with little consideration of historical relationships and interactions between related cryptocurrencies.
    In this work, we propose a generic Cross-Cryptocurrency Relationship Mining module, named \emph{C}$^2$\emph{RM}, which can effectively capture the synchronous and asynchronous impact factors between Bitcoin and related Altcoins.
    Specifically, we utilize the Dynamic Time Warping algorithm to extract the lead-lag relationship, yielding Lead-lag Variance Kernel, which will be used for aggregating the information of Altcoins to form relational impact factors.
    Comprehensive experimental results demonstrate that our \emph{C}$^2$\emph{RM} can help existing price prediction methods achieve significant performance improvement, suggesting the effectiveness of Cross-Cryptocurrency interactions on benefitting Bitcoin price prediction.

\end{abstract}
\section{Introduction} \label{sec:Introduction}
Bitcoin is the first cryptocurrency based on blockchain technology \cite{nakamoto2008Bitcoin} and is characterized by high price fluctuation \cite{aalborg2019can,balcilar2017can}. 
Since its launch in 2009, the price of Bitcoin has been in a trend of short-term fluctuation and long-term upward, and rose to nearly \$20,000 in December 2017, which consolidated its position in the mainstream market, catching the attention of investors and governments.
Meanwhile, new progress has been made in key technologies such as distributed storage, consensus mechanisms, smart contracts and encryption algorithms, laying the foundation for blockchain finance and promoting its integration into the international financial system.
In such phenomenon, more and more researchers are devoted to analyzing the trend of Bitcoin price, yielding \emph{Bitcoin price prediction}, which can help investors better deal with the changing cryptocurrency market~\cite{abay2019chainnet,liu2019bitcon,huang2019predicting}.


Due to the transparency of cryptocurrency transactions, researchers consider using transaction information such as overall trends and cyclical changes to predict cryptocurrency prices.
The most common practices~\cite{azari2019Bitcoin,mcnally2018predicting} are to use time series data of Bitcoin for price prediction, which predicts future prices using historical price information. 
Meanwhile, other work~\cite{matta2015Bitcoin,figa2019does} considers incorporating external information such as relevant policy news and Google Trends into prediction models.
In addition, several work~\cite{mudassir2020time} considers Bitcoin price prediction as a classification problem to reduce the task difficulty, i.e., transforming the regression problem of predicting specific price values into a classification problem of predicting price fluctuation.
However, existing price prediction methods suffer from several shortcomings and challenges.
First, historical price information is usually misleading in short-term forecasts, e.g. even a well-designed LSTM model with minimal error may not beat the naive strategy --- predicting that the price of the next day is equal to the previous day.
Second, external information is complex, and the impact on price fluctuation is difficult to quantify.
Third, existing price prediction models have a poor scalability or generalization, failing in transferring to new data or scenarios.

In blockchain finance market, the price fluctuation of cryptocurrencies depend not only on their historical records, but also on other economic factors and external events \cite{giudici2020cryptocurrencies}. 
In other words, the signals contained in the price fluctuation can reflect the influence of internal and external factors.
The signals are embedded in the price fluctuation of each cryptocurrency, which inspires us to use the price information of various cryptocurrencies and their interaction relationship to predict the price of another cryptocurrency.
Such practices are already present in stock price prediction.
Li et al.~\cite{li2020enhancing} considered that companies of different scales have different reaction speeds to market information, and simulates the asynchronous lead-lag relationship between stocks to predict stock trends.
Similarly, there is also a lead-lag relationship between mainstream cryptocurrencies and Altcoins~\cite{sifat2019lead}.
Most Altcoins are modeled after Bitcoin, they are based on the same consensus mechanism as it, and even completely copy its code.
Compared with Altcoins, mainstream cryptocurrencies such as Bitcoin will be more difficult to reach Nash equilibrium due to the constraints of many parties, so the price will be reflected later.

In this paper, we consider improving the performance of traditional time series prediction models for Bitcoin price prediction and propose \emph{C}$^2$\emph{RM}, a generic \textbf{C}ross-\textbf{C}ryptocurrency \textbf{R}elationship \textbf{M}ining method that can be regarded as an auxiliary module to enhance price prediction.
\emph{C}$^2$\emph{RM} can extract the synchronous and asynchronous relationship between Bitcoin and related Altcoins.
Specifically, we use Dynamic Time Warping (DTW) algorithm to extract the lead-lag relationship between Bitcoin and related Altcoins, yielding Lead-lag Variance Kernel (LVK), which will further be used for aggregating the information of Altcoins to form relational impact factors.
The relational impact factors will replace the original price series as the input of the downstream models.
Our proposed module allows for improving the performance of existing time series prediction methods for Bitcoin price prediction through cross cryptocurrency relationship extraction without adjusting them.

The main contributions of this paper are summarized as follows:
\begin{itemize}
    \item [$\bullet$] To the best of our knowledge, this is the first work to utilize the price information of various cryptocurrencies and their synchronous and asynchronous relationship to predict the price of Bitcoin. 
    \item [$\bullet$] We propose a generic cross-cryptocurrency relationship mining module, called \emph{C}$^2$\emph{RM}, which allows for extracting and aggregating the synchronous and asynchronous relationship features between Bitcoin and related Altcoins, further improving the performance of existing time series prediction methods for Bitcoin price prediction.
    \item [$\bullet$] Experimental results show the effectiveness of \emph{C}$^2$\emph{RM} module on improving the performance of existing time series prediction methods for Bitcoin price prediction.
    We study the advancement of Altcoins over Bitcoin in response to external information when Nash equilibrium is reached. 
    We also find that using short- and medium-term time series data (24 time series intervals) to predict short-term time series data (3 time series intervals) can achieve optimal investment benefits.

\end{itemize}

The remainder of this paper is organized as follows. 
In Sec.~\ref{sec:RelatedWork}, we discuss related work. 
In Sec.~\ref{sec:Methodology}, we describe the details of how our \emph{C}$^2$\emph{RM} module works. 
In Sec.~\ref{sec:Experiments}, we introduce the dataset and experimental settings, and discuss the experimental results. 
In Sec.~\ref{sec:5}, we conclude this work and present future research.

\section{Related Work}
\label{sec:RelatedWork}
Recently, the popularity of Bitcoin in the financial market has spawned a lot of research on blockchain cryptocurrencies, especially for Bitcoin price prediction, of which the related work mainly concentrates on external factors, machine learning methods and graph analytics.
\subsection{External Factors}
The earliest studies~\cite{akcora2020not} of Bitcoin aimed to trace transactions to locate the circulation of Bitcoin used for illegal activities such as money laundering and extortion.
Ladislav~\cite{kristoufek2013Bitcoin} studied the relationship between Bitcoin and search terms on Google Trends and Wikipedia, and confirmed that the fluctuation of Bitcoin prices is almost positively correlated with the number of search terms.
Bin et al.~\cite{kim2016predicting} also studied the comments on Twitter and found that comments, especially positive ones, have a very large impact on the Bitcoin price.
Karalevicius et al.~\cite{karalevicius2018using} collected and studied the database of Bitcoin-related news and blogs, and the results show that there is an interaction between media sentiment and Bitcoin price.
Gurrib~\cite{gurrib2021predicting} used linear discriminant analysis (LDA) and sentiment analysis to predict the trend of Bitcoin price fluctuation, and the results show that the LDA (SVM) model that considers both news sentiment and Bitcoin price information as input features achieves relatively good results.
These studies have shown that external information is an important factor affecting the prediction of Bitcoin price.


\subsection{Machine Learning Methods}
For Bitcoin price prediction, early work attempted to use time series data for trend prediction.
Azari~\cite{azari2019Bitcoin} used the autoregressive moving average (ARIMA) model for Bitcoin price prediction, and proved that the traditional ARIMA performs better on short-term time series data than long-term ones.
Sean et al.~\cite{mcnally2018predicting} used Bayesian optimized Recurrent Neural Network (RNN) and Long Short-Term Memory (LSTM) network to predict Bitcoin price, proving that nonlinear deep learning methods outperform ARIMA.
Stefano~\cite{cavalli2021cnn} proposed a Bitcoin trend prediction method based on one-dimensional convolutional neural network and introduced a new trading strategy, and experimental results show that the method and its strategy can more accurately predict Bitcoin trends when compared with the traditional LSTM model.


\subsection{Graph Analytics}
Akcora et al.~\cite{akcora2018forecasting} studied the impact of local topology on Bitcoin price fluctuation, and combined the ``chain'' of Bitcoin transaction network with statistical models to predict Bitcoin price.
Abay et al.~\cite{abay2019chainnet} found that the graph-related topological features show high utility in predicting Bitcoin price fluctuation.
Crowcroft et al.~\cite{crowcroft2020leveraging} considered the concept of a trusted transaction graph and proposed a set of features at a single-day time granularity. They applied the principle of autoregressive distribution lags linear regression to evaluate the intensity and duration of the changes in the features to affect the exchange rate.


    

\section{Methodology} \label{sec:Methodology}
\begin{figure}
    \centering
    \includegraphics[width=\textwidth]{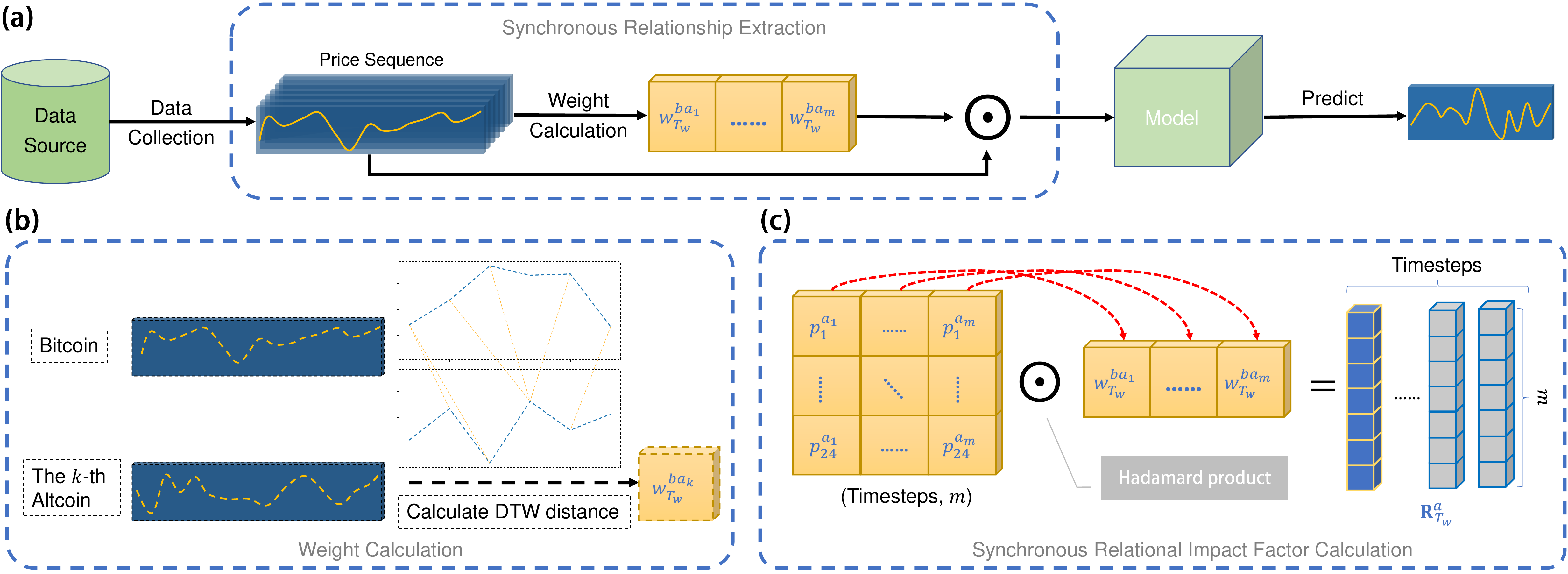}
    \caption{Illustration of synchronization method, where (a) is the overview, (b) is the process of synchronous weight calculation via DTW algorithm, and (c) is the process of aggregating Altcoin information to generate synchronous relational impact factors.} \label{fig: syn}
    \end{figure}

\begin{figure}
    \centering
    \includegraphics[width=\textwidth]{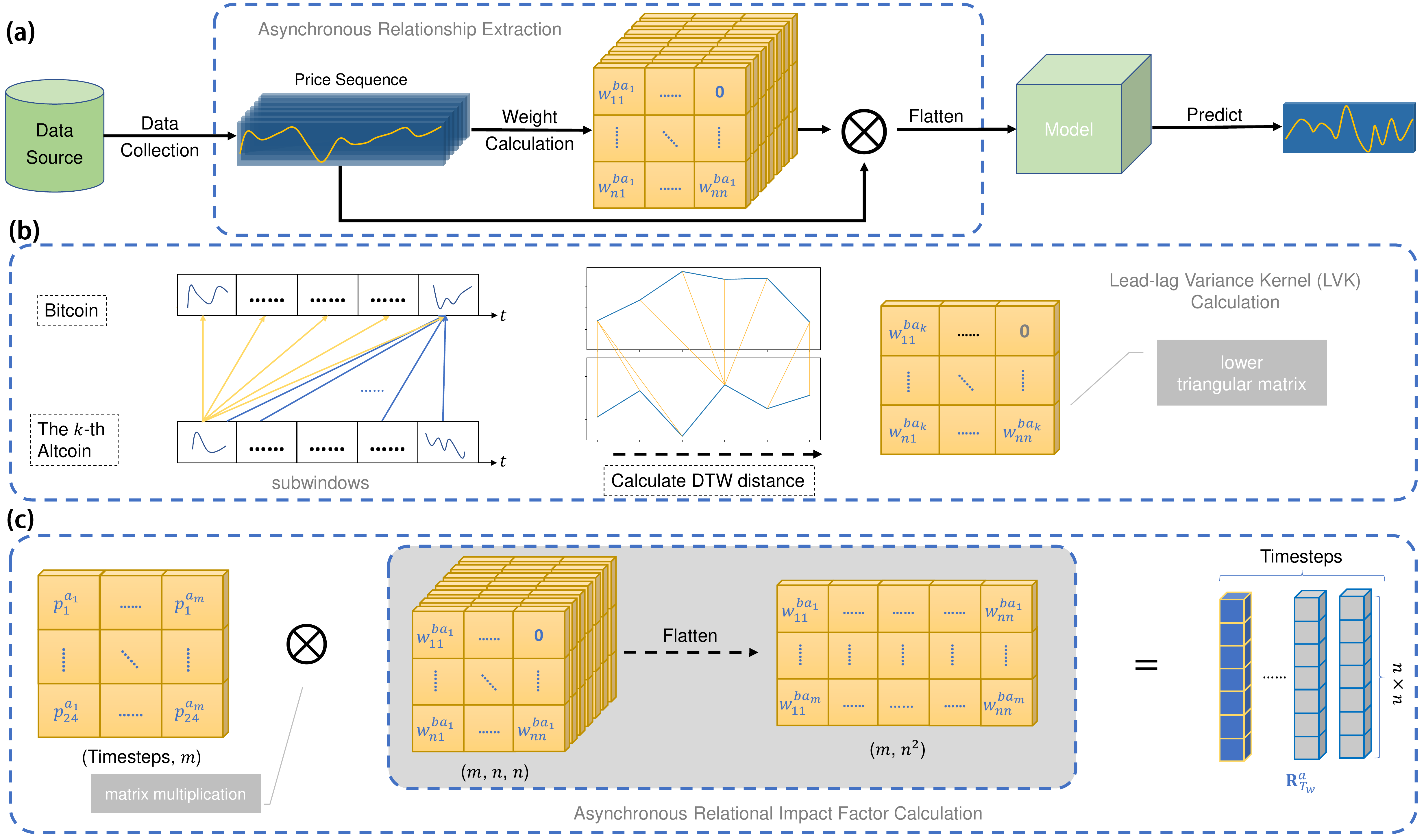}
    \caption{Illustration of asynchronous method, where (a) is the overview, (b) is the process of generating the Lead-lag Variance Kernel (LVK) via DTW algorithm, (c) is the process of aggregating Altcoin information to generate asynchronous relational impact factors.} \label{fig: asy}
\end{figure}

This section outlines our data processing approach based on the DTW algorithm for extracting information from short-term price fluctuation of seven Altcoins, as well as the overall framework for Bitcoin price prediction.
Our task is to use the historical prices of multiple Altcoins, and the fluctuation differences relative to Bitcoin within the time window, to predict the price of Bitcoin over the next few days.
To reveal information about fluctuation in Altcoins prices, we propose two methods for computing time series, in a synchronous and asynchronous manner, respectively.
Both methods are based on the DTW algorithm, and Figure~\ref{fig: syn} illustrates the synchronous method used to adjust the input sequence weights and Figure~\ref{fig: asy} explains the method of extracting asynchronous information and aggregating the input sequences.

\subsection{Data Preprocessing} \label{3.1}
\subsubsection{Normalization}
The prices of these Altcoins differ by orders of magnitude from each other, according to the market's assessment of their value.
In order to preserve their absolute impact on Bitcoin, instead of normalizing the prices of these Altcoins individually, we perform min-max normalization on them. 
We observe that the lower the price base, the greater the relative fluctuation, so lower-priced Altcoins will gain a greater DTW distance from Bitcoin over the entire time span.

\subsubsection{Time Series Segmentation}
Let $\mathcal{A} =\{a_1, a_2, \cdots, a_m\}$ represent the set of Altcoins considered in this paper, and $b$ represent the Bitcoin.
We represent the normalized prices of Altcoin $a$ at timestep $t_i$ as $p_{i}^a$.
In this paper, we make one timestep equal to 4 hours, and take the price sequences of the past 24 timesteps (96 hours) as input, and the future 3 timesteps as the label to be predicted.
For an input time window $T_w = [t_1, t_2, \cdots, t_{24}]$, the price sequences of Altcoin $a$ and Bitcoin $b$ can be represented as
\begin{equation}
    \begin{array}{l}
        \mathbf P_{T_w}^a = [p_{1}^a, p_{2}^a, \cdots, p_{24}^a], \vspace{5pt} \\ 
        \mathbf P_{T_w}^b = [p_{1}^b, p_{2}^b, \cdots, p_{24}^b].
    \end{array}
\end{equation}
The input time window size (here 24 timesteps) and the output time window size (here 3 timesteps) are a pair of hyperparameters that can be tuned.
The input time window is slid by 24 timesteps each time, that is, the input data will not be repeated.
Using more timesteps in input time window can generate LVKs with more elements, which can characterize more complex interaction relationships. 
Using more timesteps in output time window means a more difficult price prediction task.

\subsection{Lead-lag Variance Kernel} \label{3.2}
\subsubsection{DTW Distance Measure}
DTW is a method for determining the similarity of two time series that vary in different speed.
For two price curves, similar fluctuation results in a shorter DTW distance, even if there is a time shift between the two curves. 
The DTW algorithm first calculates the Euclidean distance between arbitrary two timesteps on two sequences and forms a matrix. 
Then it uses a dynamic programming algorithm to find the shortest path from the bottom right corner of the matrix to the top left corner, whose length is the sum of the matrix elements it contains.


Influenced by the entire cryptocurrency market, the prices of different currencies often fluctuate in concert. 
Our method will capture the fluctuation variance directly through DTW distance calculation.
\subsubsection{Synchronous Method}
Our synchronization method start from the DTW distance calculation between Altcoins price sequence in specific time window and Bitcoin price sequence in the same one. 
For the price sequences of Altcoin $a$ and Bitcoin $b$ in the time window $T_w$, we define the weight from Bitcoin to Altcoin as follows:
\begin{equation}
    w^{b a}_{T_w}=\text{DTW}\left(\mathbf P_{T_w}^b, \mathbf P_{T_w}^a\right).
\end{equation}
The $w^{b a}_{T_w}$ is treated as a weight, which will be multiplied to the corresponding price sequence, yielding the synchronous relational impact factor:
\begin{equation}
    \mathbf{R}_{T_w}^a = w^{b a}_{T_w} \cdot \mathbf P_{T_w}^a.
\end{equation}




\subsubsection{Asynchronous Method}
Unlike the synchronous method, asynchronous method captures the asynchronous impact of leading-lag relationship between Altcoin and Bitcoin, yielding Lead-lag Variance Kernel (LVK). 
Specifically, for two input price sequences of Bitcoin and one Altcoin, the input time windows are first divided into $n$ subwindows which contain $\frac{24}{n}$ timesteps:
\begin{equation}
    T_w = [t_{sw}^1, t_{sw}^2, \cdots, t_{sw}^n].
\end{equation}
We then calculate the DTW distance between arbitrary two subwindows between the two sequences.
Due to time causality, only the earlier subwindow will affect the later ones, the final distance of Bitcoin to a certain Altcoin form the lower triangular matrix. 
\begin{equation}
    w_{i j}^{b a}=\left\{\begin{array}{ll}
        \text{DTW}\left(\mathbf P_{t_{sw}^{i}}^b, \mathbf P_{t_{sw}^{j}}^a\right) &\  \textbf{for} \  i\geq  j  \\
        0 &\ \textbf{for} \  i<j \\
        \end{array}\right. \\
\end{equation}
where $i,j \in \{1, \cdots, n\}$, and $\mathbf W^{ba} = \{w_{i j}^{b a_{k}}\}_{m\times n\times n}$ is the LVK of Bitcoin relative to Altcoins.
The LVK will be flattened to $\bar{\mathbf{W}}^{ba} \in \mathbb{R}^{m\times n^2}$ and multiplied with the price sequences, yielding the asynchronous relational impact factor:
\begin{equation}
    \mathbf{R}_{T_w}^{ba} = \mathbf P_{T_w} \times \bar{\mathbf{W}}^{ba}.
\end{equation}


In summary, we weight and aggregate the input price sequence with element-wise multiplication on the synchronous method and matrix multiplication on the asynchronous method respectively. 
The generated relational impact factor sequences will be fed into downstream models for price prediction.

\subsection{Deep Learning Models}
As mentioned in Sec.~\ref{sec:Introduction}, deep learning models are widely used in time-dependent data analysis. 
The most popular models are RNN and its variants of LSTM and GRU. 
Compared with the typical RNN, LSTM overcomes the vanishing gradient problem\cite{hochreiter1997long}. 
GRU~\cite{cho2014learning} is similar to LSTM, but it combines the forget gate and input gate into one update gate. 
GRU has a less complex architecture than LSTM so it can be trained faster than standard RNN or LSTM. 
It has been used to forecast Bitcoin prices in the past, and is as effective as LSTM \cite{yang2021novel}, even better \cite{rizwan2019Bitcoin}.

Since our task is to predict sequences by sequences, a Seq2Seq model is necessary. 
Since we directly take the last three $1\times 1$ vectors of the hidden layer output as the result, the losses cannot be backward propagated to each RNNcell during the training process, which drives us to use bidirectional models, namely BiRNN, BiLSTM and BiGRU.
In the following, we denote the input at timestep $t$ as $\mathbf{x}_t$, which is actually the relational impact factor. 
We denote the hidden state transmitted between cells as $h_t$, and the output after the fully connected layers $f$ as $Y_{t}$. 
In this work, we use BiRNN, BiLSTM and BiGRU as downstream models to demonstrate the effectiveness of our methods.

\begin{equation}
\begin{gathered}
h_{t}^{\rightarrow}=\operatorname{Cell}\left(x_{t}, h_{t-1}^{\rightarrow}\right) \\
h_{t}^{\leftarrow}=\operatorname{Cell}\left(x_{t}, h_{t+1}^{\leftarrow}\right) \\
Y_{t}=f\left(\operatorname{concat}\left(h_{t}^{\rightarrow}, h_{t}^{\leftarrow}\right)\right)
\end{gathered}
\end{equation}

\section{Experiments} \label{sec:Experiments}
In this section, we first describe the datasets, and then present the experimental parameter settings, finally present the results and analysis.
\subsection{Dataset}
We collect the information of eight cryptocurrencies (Bitcoin and seven other Altcoins), namely Bitcoin, Ethereum, Litecoin (LTC), EOS, IOTA, Ripple (XRP), Stellar (XLM) and Cardano (ADA) on the blockchain from \textit{Binance}\footnote[1]{\url{https://data.binance.vision/}}.
For each cryptocurrency, we intercept the price from June 1, 2018 to May 1, 2020, and use 4 hours as a timestep.
Since the results of time-dependent data analysis are strongly influenced by the size of the dataset, we divided the dataset into three proportions, i.e. the ratio of the training set to the test set are $7:3$, $8:2$ and $9:1$, respectively. 




\begin{table}[htp]
    \setlength{\tabcolsep}{4mm}
    \centering\
    \renewcommand{\arraystretch}{1}
    \caption{Data description of cryptocurrencies used in this paper.}
    \resizebox{\textwidth}{!}{%
    \begin{tabular}{crrc} 
    \hline\hline
    \textbf{Cryptocurrency} & \textbf{Min-price} (\$) & \textbf{Max-price} (\$) & \textbf{Time Interval}                                                                \\ 
    \cline{1-4}
    Bitcoin          & 3156.26                 & 13960.76  & \multirow{8}{*}{\begin{tabular}[c]{@{}c@{}}June 1, 2018\\$\sim$\\May 1, 2020\end{tabular}}  \\
    Ethereum         & 81.79                   & 628.81    &                                                                    \\
    Litecoin         & 22.32                   & 145.90    &                                                                    \\
    EOS              & 1.36                    & 15.68     &                                                                    \\
    IOTA             & 0.05                    & 2.01      &                                                                    \\
    Ripple           & 0.10                    & 0.80      &                                                                    \\
    Stellar          & 0.03                    & 0.35      &                                                                    \\
    Cardano          & 0.02                    & 0.24      &                                                                    \\
    \hline\hline
    \end{tabular}}
    \label{tb: data}
\end{table}

\subsection{Baselines}
To illustrate the effectiveness of our \emph{C}$^2$\emph{RM} module, we combine it with three RNN-based models, and compare with machine learning model and MLPs.
The parameter settings are summarized in Table~\ref{tb: para}.
\subsubsection{GBRT}
The basic settings and hyperparameters of the baselines remain the same as in our proposed approach, except for the machine learning method. 
GBRT is a non-parametric statistical learning technique for regression prediction. 
Since \cite{elsayed2021we} argued that GBRT is no worse than DNN for time series prediction problems, we apply it to our task as well.

\subsubsection{RNNs}
We employ three of the most common time series prediction models, namely standard \textbf{RNN}, \textbf{LSTM} and \textbf{GRU}, all involving a bi-directional structure. 
$ReLU$ is employed as the activation function in the cells of standard RNN. To control the variables, we use a uniform hyperparameter settings for a fair comparison.

\subsubsection{MLPs} 
We first present a \textbf{SmartMLP} method, in which we directly input ``\textbf{Input window} $\times$ \textbf{Number of Coin Types}'' into fully connected layers and output a vector with length equal to output window.
We also take the same input and output a scalar, and broadcast it to a vector with length equal to output window, which is the \textbf{NaiveMLP} method.

\begin{table}[htp]
    \setlength{\tabcolsep}{3mm}
    \centering
    \renewcommand{\arraystretch}{1}
    \caption{Hyperparameter settings.}
    \label{tb: para}
    \resizebox{\textwidth}{!}{%
    \begin{tabular}{cc|cc} 
    \hline\hline
    \multicolumn{2}{c|}{BiRNNs}           & \multicolumn{2}{c}{GBRT}         \\ 
    \hline
    Shape of LVK             & 4$\times$4 & Number of estimators     & 200   \\
    Hidden dimension         & 32         & Max depth                & 6     \\
    Layers (both directions) & 2          & Min child weight         & 1     \\
    Input window size        & 24         & L1 regularization factor & 0.9   \\
    Output window size       & 3          & L2 regularization factor & 1     \\
    Input dimension          & 7          & Learning rate            & 0.01  \\
    Epochs                   & 2000       &                          &       \\
    Learning rate            & 0.01       &                          &       \\
    \hline\hline
    \end{tabular}}
\end{table}
\begin{table}
    \centering
    \caption{Summary of performance on Bitcoin price prediction under different data split schemes in terms of MSE.
    The highest performance is marked with boldface; the highest performance of baselines is underline.}
    \renewcommand\arraystretch{1.2}
    \resizebox{\textwidth}{30mm}{
    \setlength{\tabcolsep}{5mm}{
    \begin{tabular}{lcccc} 
    \hline\hline
    \multirow{2}{*}{\textbf{Methods}}    & \multicolumn{3}{c}{\textbf{Data Split} ($MSE \times 10^{-3}$)}                   & \multicolumn{1}{l}{\multirow{2}{*}{\textbf{Mean}($MSE \times 10^{-3}$)}}  \\ 
    \cline{2-4}
                                        & 7:3            & 8:2            & 9:1            & \multicolumn{1}{l}{}                       \\ 
    \hline                        
    GBRT                             & 8.682                      & 3.668                      & 2.513                      & 4.954           \\
    \hdashline                                          
    Naive-MLP                        & 4.137                      & 4.975                      & 1.017                      & 3.376           \\
    Smart-MLP                        & 4.832                      & 5.597                      & \underline{0.885}          & 3.771           \\ 
    \hdashline    
    BiRNN                            & \underline{2.002}          & \underline{2.498}          & 0.959                      & \underline{1.820}           \\
    BiLSTM                           & 3.588                      & 3.346                      & 0.916                      & 2.617           \\
    BiGRU                            & 2.613                      & 2.541                      & 1.439                      & 2.198           \\
    \hline 
    BiRNN-syn\emph{C}$^2$\emph{RM}   & 1.323                      & 0.580                      & 1.036                      & 0.980           \\
    BiLSTM-syn\emph{C}$^2$\emph{RM}  & 1.657                      & 0.637                      & 1.014                      & 1.103           \\                            BiGRU-syn\emph{C}$^2$\emph{RM}   & 1.127                      & 0.823                      & 1.143                      & 1.031           \\ 
    \hdashline
    BiRNN-asyn\emph{C}$^2$\emph{RM}  & \textbf{0.837}             & 0.445                      & 0.824                      & \textbf{0.702}  \\
    BiLSTM-asyn\emph{C}$^2$\emph{RM} & 0.924                      & 0.494                      & \textbf{0.757}             & 0.725           \\
    BiGRU-asyn\emph{C}$^2$\emph{RM}  & 1.080                      & \textbf{0.420}             & 0.779                      & 0.760           \\
\hline\hline
\end{tabular}}}
\label{tb: result}
\end{table}
    
\subsection{Evaluation Metric}
Assume that the predictive value is
\begin{equation}
\hat{\mathbf{y}}=\left\{\hat{y}_{1}, \hat{y}_{2}, \cdots, \hat{y_{n}}\right\},
\end{equation} 
the ground truth is \begin{equation}
\mathbf{y}=\left\{y_{1}, y_{2}, \cdots, y_{n}\right\}.
\end{equation}
\subsubsection{MSE(Mean Square Error)} is a common metric used in regression problems to assess the prediction results. 
It is defined as the average squared difference between the predictive values and the actual values:
\begin{equation}
MSE=\frac{1}{n} \sum_{i=1}^{n}\left(\hat{y}_{i}-y_{i}\right)^{2}.
\end{equation}


\subsection{Results and Analysis}
This section analyses the results of our experiments, including a comparison with baselines, a comparison of synchronous and asynchronous methods, and additional experiments to illustrate the generalization of the asynchronous method.

\begin{figure}
    \centering
    \includegraphics[width=\textwidth]{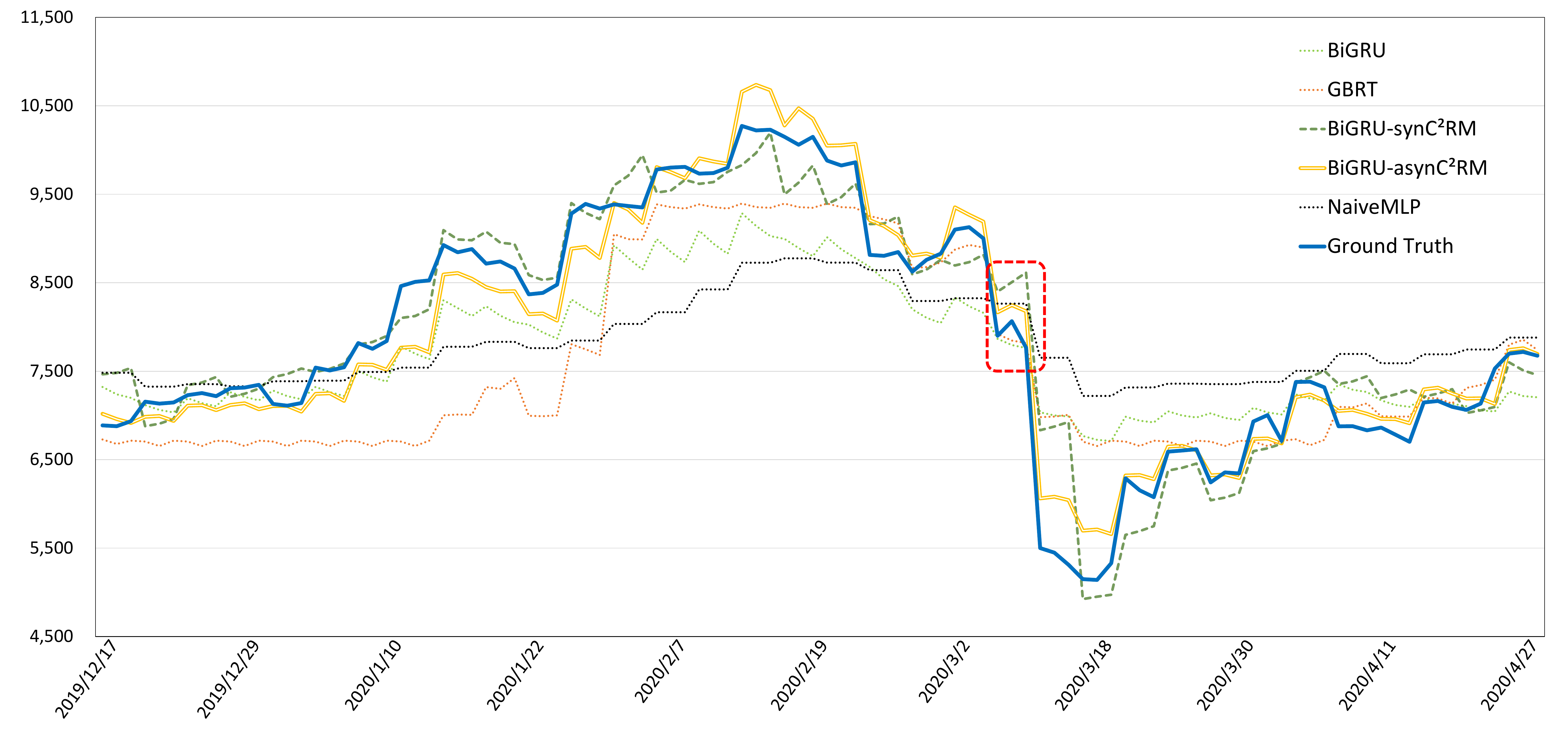}
    \caption{The results of the 5 price predicting methods as well as the ground truth. The units on the vertical axis are the actual price against the US dollar. The horizontal axis is the actual date corresponding to every 3 consecutive time points at an interval of 24 points.} 
    \label{fig: result}
\end{figure}  

\subsubsection{Enhancement for Bitcoin Price Prediction}
Table~\ref{tb: result} reports the results of performance comparison between raw methods and their enhanced version (with \emph{C}$^2$\emph{RM}) with different data split schemes, from which we observe that there is a significant boost in prediction performance across all RNN-based methods, suggesting the effectiveness of our \emph{C}$^2$\emph{RM} module on helping existing price prediction methods achieve performance improvement.

Specifically, the machine learning method (GBRT) obtains relative lower performance rankings, indicating the limited expressiveness of the shallow features learnt by GBRT model.
The MLP-based methods significantly outperform GBRT in most cases, and gain 23.88\% and 31.85\% average relative improvements, suggesting that the simple deep models and naive prediction strategies can provide certain price prediction guidance.
The RNN-based methods achieve the best prediction performance across all baselines, indicating that such time series prediction models can effectively capture the time dependence of price fluctuation. 
Finally, these RNN-based methods (BiRNN, BiLSTM and BiGRU) combined with the proposed \emph{C}$^2$\emph{RM} module obtain higher prediction performance, yielding 45.15\%, 57.85\% and 46.91\% average relative improvements respectively in terms of synchronous method, and 61.43\%, 72.30\% and 65.42\% average relative improvements respectively in terms of asynchronous method, when compared with these RNN-based methods that accept raw price sequence input.
Such phenomenon demonstrates that our \emph{C}$^2$\emph{RM} module can effectively extract the impact factors of related Altcoins on Bitcoin, further helping to predict the price of Bitcoin more accurately.
Furthermore, we also observe that asyn\emph{C}$^2$\emph{RM} totally beat syn\emph{C}$^2$\emph{RM}, indicating that the lead-lag relationship between Altcoins and Bitcoin benefits more than the synchronous relationship.

\begin{figure}
    \centering
    \includegraphics[width=0.8\textwidth]{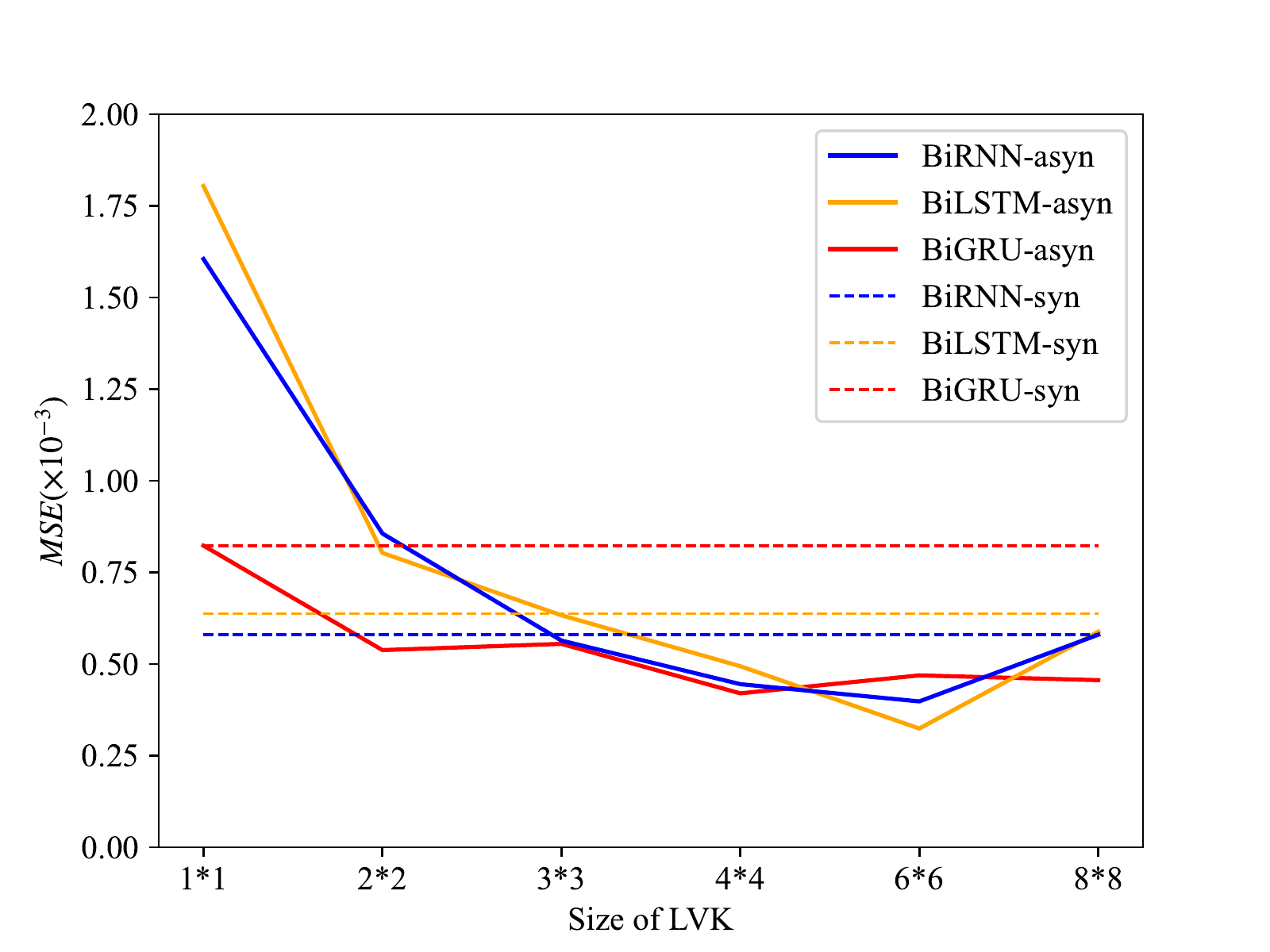}
    \caption{Effect of LVK size on the performance of BiRNNs}
    \label{fig: size}
\end{figure}
\subsubsection{Overall Trend Prediction}
We further analyze the performance of different methods in overall trend prediction, as illustrated in Fig.~\ref{fig: result}, from which we observe that all methods are able to predict the overall trend of Bitcoin price while BiGRU-asyn\emph{C}$^2$\emph{RM} represented by the hollow solid line performs best.
Specifically, the price curve predicted by BiGRU-asyn\emph{C}$^2$\emph{RM} is closest to the ground truth, and BiGRU-syn\emph{C}$^2$\emph{RM} is the second closest.
Such phenomenon suggests that our \emph{C}$^2$\emph{RM} module can effectively capture the price fluctuation and the asynchronous lead-lag relationship benefits more.

Furthermore, as mentioned in Sec.~\ref{sec:Introduction}, even for the task of predicting one point in the future, RNNs simply repeat the last day of the input window in a highly weighted way so that they cannot beat NAIVE methods that repeat the last point exactly.
Therefore, it makes more sense to focus on the last two points in a three-point prediction, or the trend in a single point prediction, because the values of the last two points are not inferred from the first point.
In the three-point prediction framed by the dashed rectangle, our \emph{C}$^2$\emph{RM} also demonstrates its ability to predict three-point trends while other methods fail.
In this output window, the actual price shows a trend of first rising and then falling, which is difficult to predict. 
Our asyn\emph{C}$^2$\emph{RM} captures such pattern while other methods unsurprisingly employ the naive strategy of repeating the previous trend, i.e. maintaining two consecutive up or down trends.


\subsubsection{Impact of LVK Size}
We further investigate the impact of the most important hyperparameter --- the size of LVK ($n \times n$) in the asyn\emph{C}$^2$\emph{RM} method. 
Specifically, we vary the number of subwindows $n$ in ${1,2,3,4,6,8}$ and fix other hyperparameters.
We present the evaluation results in Fig.~\ref{fig: size}, from which we observe that larger LVK (with more subsections in input sequence) benefits price prediction more. But when the size arrives at a certain level, improvements become slighter.
We speculate that larger LVK can extract more complex and detailed leading-lag relationship between Bitcoin and related Altcoins, yielding more powerful generalization.

\section{Conclusion} \label{sec:5}
In this paper, we propose two generic modules for Bitcoin price prediction, including syn\emph{C}$^2$\emph{RM} for extracting synchronous relationship and asyn\emph{C}$^2$\emph{RM} for extracting asynchronous lead-lag relationship. 
Experiments demonstrate the effectiveness of our methods in enhancing the existing Bitcoin price prediction methods.
We believe that the Lead-lag Variance Kernel in asyn\emph{C}$^2$\emph{RM} successfully extracts the difference in short-term fluctuation between the historical prices of Altcoins and Bitcoin and the impact on future price fluctuation. 
Moreover, we also investigate the generalization and effectiveness of LVK in terms of size variation.


\subsubsection*{Acknowledgments.} This work was partially supported by the National Key R\&D Program of China under Grant 2020YFB1006104, by the Key R\&D Programs of Zhejiang under Grants 2022C01018 and 2021C01117, by the National Natural Science Foundation of China under Grant 61973273, and by the Zhejiang Provincial Natural Science Foundation of China under Grant LR19F030001.
%
%
%
\bibliographystyle{splncs04_}
\bibliography{ref}

\end{document}